\documentclass[aps,prd,twocolumn,superscriptaddress,nofootinbib]{revtex4-1}

\usepackage{latexsym}
\usepackage{amsmath}
\usepackage{amssymb}
\usepackage{amsfonts}
\usepackage{bm}

\usepackage{color}

\usepackage{supertabular} 
\usepackage{placeins}
\usepackage{epsfig}
\usepackage{graphicx}

\begin{document}

%\preprint{}

\title{Hidden-Bottom Pentaquarks}


\author{Gang Yang }
\email[]{ygz0788a@sina.com}
%\thanks{}
\affiliation{Department of Physics and State Key Laboratory of Low-Dimensional Quantum Physics, \\ Tsinghua University, Beijing 100084, P. R. China}

\author{Jialun Ping}
\email[]{jlping@njnu.edu.cn}
\affiliation{Department of Physics and Jiangsu Key Laboratory for Numerical Simulation of Large Scale Complex Systems, Nanjing Normal University, Nanjing 210023, P. R. China}

\author{Jorge Segovia}
\email[]{jsegovia@ifae.es}
\affiliation{Institut de F\'isica d'Altes Energies (IFAE) and Barcelona Institute of Science and Technology (BIST), \\ Universitat Aut\`onoma de Barcelona, E-08193 Bellaterra (Barcelona), Spain}


\date{\today}

\begin{abstract}
The LHCb Collaboration has recently reported strong evidences of the existence of pentaquark states in the hidden-charm baryon sector, the so-called $P_c(4380)^+$ and $P_c(4450)^+$ signals. Five-quark bound states in the hidden-charm sector were explored by us using, for the quark-quark interaction, a chiral quark model which successfully explains meson and baryon phenomenology, from the light to the heavy quark sector. We extend herein such study but to the hidden-bottom pentaquark sector, analyzing possible bound-states with spin-parity quantum numbers $J^P=\frac12^\pm$, $\frac32^\pm$ and $\frac52^\pm$, and in the $\frac12$ and $\frac32$ isospin sectors. We do not find positive parity hidden-bottom pentaquark states; however, several candidates with negative parity are found with dominant baryon-meson structures $\Sigma_b^{(\ast)}\bar{B}^{(\ast)}$.
The calculated distances among any pair of quarks within the bound-state reflect that molecular-type bound-states are favored when only color-singlet configurations are considered in the coupled-channels calculation whereas compact pentaquarks, which are also deeply bound, can be found when hidden-color configurations are added.
Finally, our findings resemble the ones found in the hidden-charm sector but, as expected, we find in the hidden-bottom sector larger binding energies and bigger contributions of the hidden-color configurations.
\end{abstract}

\pacs{
12.38.-t \and % Quantum Chromodynamics
12.39.-x \and % Potential Models
14.20.-c \and % Properties of Baryons
14.20.Pt      % Exotic Baryons
}
\keywords{
Quantum Chromodynamics \and
Quark models           \and
Properties of Baryons  \and
Exotic Baryons
}

\maketitle

%%%%%%%%%%%%%%%%%%%%%%%%%%%%%%%%%%%%%%%%%%%%%%%%%%%%%%%%%%%%%%%%%%%%%%%%%%%%%%%%%%%%%%%%%%

\section{Introduction}

After decades of experimental and theoretical studies of hadrons, the conventional picture of mesons and baryons as, respectively, quark-antiquark and $3$-quark bound states is being left behind. On one hand, Quantum Chromodynamics (QCD), the non-Abelian quantum field theory of the strong interactions, does not prevent to have exotic hadrons such as glueballs, quark-gluon hybrids and multiquark systems. On the other hand, more than two dozens of nontraditional charmonium- and bottomonium-like states, the so-called XYZ mesons, have been observed in the last 15 years at B-factories (BaBar, Belle and CLEO), $\tau$-charm facilities (CLEO-c and BESIII) and also proton-(anti)proton colliders (CDF, D0, LHCb, ATLAS and CMS). 

In $2015$, the LHCb Collaboration observed two hidden-charm pentaquark states in the $J/\psi p$ invariant mass spectrum of the $\Lambda^{0}_{b} \rightarrow J/\psi K^{-}p$ decay~\cite{Aaij:2015tga}. One is $P_{c}(4380)^+$ with a mass of $(4380\pm8\pm29)\,\text{MeV}$ and a width of $(205\pm18\pm86)\,\text{MeV}$, and another is $P_{c}(4450)^+$ with a mass of $(4449.8\pm1.7\pm2.5)\,\text{MeV}$ and a width of $(39\pm5\pm19)\,\text{MeV}$. The preferred $J^{P}$ assignments are of opposite parity, with one state having spin $\frac{3}{2}$ and the other $\frac{5}{2}$.

The discovery of the $P_c(4380)^+$ and $P_c(4450)^+$ has triggered a strong theoretical interest on multiquark systems. The interested reader is directed to the recent review~\cite{Chen:2016qju} in order to have a global picture of the current progress; however, one can highlight those theoretical studies of the $P_c(4380)^+$ and $P_c(4450)^+$ in which different kind of quark arrangements are used such as diquark-triquark~\cite{Wang:2016dzu, Zhu:2015bba, Lebed:2015tna}, diquark-diquark-antiquark~\cite{Wang:2016dzu, Anisovich:2015cia, Maiani:2015vwa, Ghosh:2015ksa, Wang:2015ava, Wang:2015epa, Wang:2015ixb}, and meson-baryon molecule~\cite{Wang:2016dzu, Roca:2015dva, Chen:2015loa, Huang:2015uda, Meissner:2015mza, Xiao:2015fia, He:2015cea, Chen:2015moa, Chen:2016heh, Yamaguchi:2016ote, He:2016pfa, Ortega:2016syt, Azizi:2017bgs, Anwar:2018bpu}. It is also noteworthy that some recent investigations have considered other possible physical mechanisms as the origin of the experimental signals like kinematic effects and triangle singularities~\cite{Guo:2015umn, Mikhasenko:2015vca, Liu:2015fea, Liu:2016dli, Bayar:2016ftu}.

The observation of hadrons containing valence $c$-quarks is historically followed by the identification of similar structures with $b$-quark content. Therefore, it is natural to expect a subsequent observation of the bottom analogues of the $P_c(4380)^+$ and $P_c(4450)^+$ resonances. In fact, the LHCb Collaboration has recently made an attempt to search for pentaquark states containing a single $b$-quark, that decays weakly via the $b\to c\overline{c}s$ transition, in the final states $J/\psi K^+\pi^- p$, $J/\psi K^- \pi^- p$, $J/\psi K^- \pi^+ p$, and $J/\psi \phi p$~\cite{Aaij:2017jgf}; and thus reports about similar explorations in the hidden-bottom pentaquark sector should be expected in the near future.

Theoretical investigations of the spectrum of hidden-bottom pentaquarks as well as their electromagnetic, strong and weak decays help in the experimental hunt mentioned above. In addition to this, further theoretical studies supply complementary information on the internal structure and inter-quark interactions of pentaquarks with heavy quark content. In Ref.~\cite{Shimizu:2016rrd}, besides the $P_c(4380)^+$ state, the possible existence of hidden-bottom pentaquarks with a mass around $11.08-11.11\,\text{GeV}$ and quantum numbers $J^{P}={3/2}^-$ was emphasized; it was also indicated that there may exist some loosely-bound molecular-type pentaquarks in other heavy quark sectors. See also Refs.~\cite{Wu:2010jy, Wu:2010rv} for more information on the properties of the charmed and bottom pentaquark states using the coupled-channel unitary approach as well as Refs. \cite{Guo:2015umn, Liu:2015fea, Guo:2016bkl, Bayar:2016ftu} for illuminating discussions on the structure of pentaquarks and their possible relation with triangle singularities.

We study herein, within a chiral quark model formalism, the possibility of having pentaquark bound-states in the hidden-bottom sector with quantum numbers $J^P=\frac12^\pm$, $\frac32^\pm$ and $\frac52^\pm$, and in the $\frac12$ and $\frac32$ isospin sectors. This work is a natural extension of the analysis performed in Ref.~\cite{Yang:2015bmv} in which similar structures were studied but in the hidden-charm sector. In Ref.~\cite{Yang:2015bmv}, the $P_c(4380)^+$ was suggested to be a bound state of $\Sigma_c^\ast\bar{D}$ with quantum numbers $J^P=\frac{3}{2}^-$ whereas the nature of the $P_c(4450)^+$ structure was not clearly established because, despite of having a couple of possible candidates attending to the agreement between theoretical and experimental masses, there was an inconsistency between the parity of the state determined experimentally and those predicted theoretically. Further pentaquark bound-states which contain dominant $\Sigma_c\bar{D}$ and $\Sigma_c^\ast\bar{D}^\ast$ Fock-state components were also found in the region about $4.3-4.5\,\text{GeV}$. 

All the details about our computational framework will be described later but let us sketch here some of its main features. Our chiral quark model (ChQM) is based on the fact that chiral symmetry is spontaneously broken in QCD and, among other consequences, it provides a constituent quark mass to the light quarks. To restore the chiral symmetry in the QCD Lagrangian, Goldstone-boson exchange interactions appear between the light quarks. This fact is encoded in a phenomenological potential which already contains the perturbative one-gluon exchange (OGE) interaction and a nonperturbative linear-screened confining term.\footnote{The interested reader is referred to Refs.~\cite{Valcarce:2005em, Segovia:2013wma} for detailed reviews on the naive quark model in which this work is based.} It is worth to note that chiral symmetry is explicitly broken in the  heavy quark sector and this translates in our formalism to the fact that the interaction terms between light-light, light-heavy and heavy-heavy quarks are not the same, i.e. while Goldstone-boson exchanges are considered when the two quarks are light, they do not appear in the other two configurations: light-heavy and heavy-heavy; however, the one-gluon exchange and confining potentials are flavor blindness.

The five-body bound state problem is solved by means of the Gau\ss ian expansion method (GEM)~\cite{Hiyama:2003cu} which has been demonstrated to be as accurate as a Faddeev calculation (see, for instance, Figs.~15 and~16 of Ref.~\cite{Hiyama:2003cu}). As it is well know, the quark model parameters are crucial in order to describe particular physical observables. We have used values that have been fitted before through hadron~\cite{Valcarce:1995dm, Vijande:2004he, Segovia:2008zza, Segovia:2008zz, Ortega:2016hde, Yang:2017xpp}, hadron-hadron ~\cite{Fernandez:1993hx, Valcarce:1994nr, Ortega:2009hj, Ortega:2016mms, Ortega:2016pgg} and multiquark~\cite{Vijande:2006jf, Yang:2015bmv, Yang:2017rpg} phenomenology.

The structure of the present manuscript is organized in the following way. In Sec.~\ref{sec:model} the ChQM, pentaquark wave-functions and GEM are briefly presented and discussed. Section~\ref{sec:results} is devoted to the analysis and discussion on the obtained results. We summarize and give some prospects in Sec.~\ref{sec:summary}.

%%%%%%%%%%%%%%%%%%%%%%%%%%%%%%%%%%%%%%%%%%%%%%%%%%%%%%%%%%%%%%%%%%%%%%%%%%%%%%%%%%%%%%%%%%

\section{Theoretical framework}
\label{sec:model}

Although Lattice QCD (LQCD) has made an impressive progress on understanding multiquark systems~\cite{Alexandrou:2001ip, Okiharu:2004wy} and the hadron-hadron interaction~\cite{Prelovsek:2014swa, Lang:2014yfa, Briceno:2017max}, the QCD-inspired quark models are still the main tool to shed some light on the nature of the multiquark candidates observed by experimentalists.

The general form of our five-body Hamiltonian is given by~\cite{Yang:2015bmv}
\begin{equation}
H = \sum_{i=1}^{5}\left( m_i+\frac{\vec{p\,}^2_i}{2m_i}\right) - T_{\text{CM}} + \sum_{j>i=1}^{5} V(\vec{r}_{ij}) \,,
\label{eq:Hamiltonian}
\end{equation}
where $T_{\text{CM}}$ is the center-of-mass kinetic energy and the two-body potential
\begin{equation}
V(\vec{r}_{ij}) = V_{\text{CON}}(\vec{r}_{ij}) + V_{\text{OGE}}(\vec{r}_{ij}) + V_{\chi}(\vec{r}_{ij}) \,,
\end{equation}
includes the color-confining, one-gluon exchange and Goldstone-boson exchange interactions. Note herein that the potential could contain central, spin-spin, spin-orbit and tensor contributions; only the first two will be considered attending the goal of the present manuscript and for clarity in our discussion.

Color confinement should be encoded in the non-Abelian character of QCD. Studies of lattice-regularized QCD have demonstrated that multi-gluon exchanges produce an attractive linearly rising potential proportional to the distance between infinite-heavy quarks~\cite{Bali:2005fu}. However, the spontaneous creation of light-quark pairs from the QCD vacuum may give rise at the same scale to a breakup of the created color flux-tube~\cite{Bali:2005fu}. We have tried to mimic these two phenomenological observations by the expression:
\begin{equation}
V_{\text{CON}}(\vec{r}_{ij}\,)=\left[-a_{c}(1-e^{-\mu_{c}r_{ij}})+\Delta \right] 
(\vec{\lambda}_{i}^{c}\cdot\vec{\lambda}_{j}^{c}) \,,
\label{eq:conf}
\end{equation}
where $a_{c}$ and $\mu_{c}$ are model parameters, and the SU(3) color Gell-Mann matrices are denoted as $\lambda^c$. One can see in Eq.~\eqref{eq:conf} that the potential is linear at short inter-quark distances with an effective confinement strength $\sigma = -a_{c} \, \mu_{c} \, (\vec{\lambda}^{c}_{i}\cdot \vec{\lambda}^{c}_{j})$, while it becomes constant at large distances. 

The one-gluon exchange potential is given by
\begin{align}
&
V_{\text{OGE}}(\vec{r}_{ij}) = \frac{1}{4} \alpha_{s} (\vec{\lambda}_{i}^{c}\cdot
\vec{\lambda}_{j}^{c}) \Bigg[\frac{1}{r_{ij}} \nonumber \\ 
&
\hspace*{1.60cm} - \frac{1}{6m_{i}m_{j}} (\vec{\sigma}_{i}\cdot\vec{\sigma}_{j}) 
\frac{e^{-r_{ij}/r_{0}(\mu)}}{r_{ij}r_{0}^{2}(\mu)} \Bigg] \,,
\end{align}
where $m_{i}$ is the quark mass and the Pauli matrices are denoted by $\vec{\sigma}$. The contact term of the central potential has been regularized as
\begin{equation}
\delta(\vec{r}_{ij})\sim\frac{1}{4\pi r_{0}^{2}}\frac{e^{-r_{ij}/r_{0}}}{r_{ij}} \,,
\end{equation}
with $r_{0}(\mu_{ij})=\hat{r}_{0}/\mu_{ij}$ a regulator that depends on $\mu_{ij}$, the reduced mass of the quark--(anti-)quark pair.

The wide energy range needed to provide a consistent description of mesons and baryons from light to heavy quark sectors requires an effective scale-dependent strong coupling constant. We use the frozen coupling constant of, for instance, Ref.~\cite{Segovia:2013wma}
\begin{equation}
\alpha_{s}(\mu_{ij})=\frac{\alpha_{0}}{\ln\left(\frac{\mu_{ij}^{2}+\mu_{0}^{2}}{\Lambda_{0}^{2}} \right)} \,,
\end{equation}
in which $\alpha_{0}$, $\mu_{0}$ and $\Lambda_{0}$ are parameters of the model.

The central terms of the chiral quark--(anti-)quark interaction can be written as
\begin{align}
&
V_{\pi}\left( \vec{r}_{ij} \right) = \frac{g_{ch}^{2}}{4\pi}
\frac{m_{\pi}^2}{12m_{i}m_{j}} \frac{\Lambda_{\pi}^{2}}{\Lambda_{\pi}^{2}-m_{\pi}
^{2}}m_{\pi} \Bigg[ Y(m_{\pi}r_{ij}) \nonumber \\
&
\hspace*{1.20cm} - \frac{\Lambda_{\pi}^{3}}{m_{\pi}^{3}}
Y(\Lambda_{\pi}r_{ij}) \bigg] (\vec{\sigma}_{i}\cdot\vec{\sigma}_{j})\sum_{a=1}^{3}(\lambda_{i}^{a}
\cdot\lambda_{j}^{a}) \,, \\
& 
V_{\sigma}\left( \vec{r}_{ij} \right) = - \frac{g_{ch}^{2}}{4\pi}
\frac{\Lambda_{\sigma}^{2}}{\Lambda_{\sigma}^{2}-m_{\sigma}^{2}}m_{\sigma} \Bigg[
Y(m_{\sigma}r_{ij}) \nonumber \\
&
\hspace*{1.20cm} - \frac{\Lambda_{\sigma}}{m_{\sigma}}Y(\Lambda_{\sigma}r_{ij})
\Bigg] \,, \\
& 
V_{K}\left( \vec{r}_{ij} \right)= \frac{g_{ch}^{2}}{4\pi}
\frac{m_{K}^2}{12m_{i}m_{j}} \frac{\Lambda_{K}^{2}}{\Lambda_{K}^{2}-m_{K}^{2}}m_{
K} \Bigg[ Y(m_{K}r_{ij}) \nonumber \\
&
\hspace*{1.20cm} -\frac{\Lambda_{K}^{3}}{m_{K}^{3}}Y(\Lambda_{K}r_{ij})
\Bigg] (\vec{\sigma}_{i}\cdot\vec{\sigma}_{j})\sum_{a=4}^{7}(\lambda_{i}^{a}
\cdot\lambda_{j}^{a}) \,, \\
& 
V_{\eta}\left( \vec{r}_{ij} \right) = \frac{g_{ch}^{2}}{4\pi}
\frac{m_{\eta}^2}{12m_{i}m_{j}} \frac{\Lambda_{\eta}^{2}}{\Lambda_{\eta}^{2}-m_{
\eta}^{2}}m_{\eta} \Bigg[ Y(m_{\eta}r_{ij}) \nonumber \\
&
\hspace*{1.20cm} -\frac{\Lambda_{\eta}^{3}}{m_{\eta}^{3}
}Y(\Lambda_{\eta}r_{ij}) \Bigg] (\vec{\sigma}_{i}\cdot\vec{\sigma}_{j})
\Big[\cos\theta_{p} \left(\lambda_{i}^{8}\cdot\lambda_{j}^{8}
\right) \nonumber \\
&
\hspace*{1.20cm} -\sin\theta_{p} \Big] \,,
\end{align}
where $Y(x)$ is the standard Yukawa function defined by $Y(x)=e^{-x}/x$. We consider the physical $\eta$ meson instead of the octet one and so we introduce the angle $\theta_p$. The $\lambda^{a}$ are the SU(3) flavor Gell-Mann matrices. Taken from their experimental values, $m_{\pi}$, $m_{K}$ and $m_{\eta}$ are the masses of the SU(3) Goldstone bosons. The value of $m_{\sigma}$ is determined through the PCAC relation $m_{\sigma}^{2}\simeq m_{\pi}^{2}+4m_{u,d}^{2}$~\cite{Scadron:1982eg}. Finally, the chiral coupling constant, $g_{ch}$, is determined from the $\pi NN$ coupling constant through
\begin{equation}
\frac{g_{ch}^{2}}{4\pi}=\frac{9}{25}\frac{g_{\pi NN}^{2}}{4\pi} \frac{m_{u,d}^{2}}{m_{N}^2} \,,
\end{equation}
which assumes that flavor SU(3) is an exact symmetry only broken by the different mass of the strange quark.

The model parameters have been fixed in advance reproducing hadron~\cite{Valcarce:1995dm, Vijande:2004he, Segovia:2008zza, Segovia:2008zz, Ortega:2016hde, Yang:2017xpp}, hadron-hadron ~\cite{Fernandez:1993hx, Valcarce:1994nr, Ortega:2009hj, Ortega:2016mms, Ortega:2016pgg} and multiquark~\cite{Vijande:2006jf, Yang:2015bmv, Yang:2017rpg} phenomenology. For clarity, the ones involved in this calculation are listed in Table~\ref{model}. They were used in Ref.~\cite{Yang:2015bmv} to study possible hidden-charm pentaquark bound-states with quantum numbers $IJ^P=\frac{1}{2}\left(\frac{1}{2}\right)^\pm$, $\frac{1}{2}\left(\frac{3}{2}\right)^\pm$ and $\frac{1}{2}\left(\frac{5}{2}\right)^\pm$; moreover, their properties were compared with those associated with the hidden-charm pentaquark signals observed by the LHCb Collaboration in Ref.~\cite{Aaij:2015tga}.

\begin{table}[!t]
\caption{\label{model} Quark model parameters.}
\begin{ruledtabular}
\begin{tabular}{cccc}
Quark masses     & $m_u=m_d$ (MeV) &  313 \\
                 & $m_b$ (MeV)     & 5100 \\[2ex]
Goldstone bosons & $\Lambda_\pi=\Lambda_\sigma~$ (fm$^{-1}$) &   4.20 \\
                 & $\Lambda_\eta$ (fm$^{-1}$)     &   5.20 \\
                 & $g^2_{ch}/(4\pi)$                         &   0.54 \\
                 & $\theta_P(^\circ)$                        & -15 \\[2ex]
Confinement      & $a_c$ (MeV)         & 430\\
                 & $\mu_c$ (fm$^{-1})$ &   0.70\\
                 & $\Delta$ (MeV)      & 181.10 \\[2ex]
                 & $\alpha_0$              & 2.118 \\
                 & $\Lambda_0~$(fm$^{-1}$) & 0.113 \\
OGE              & $\mu_0~$(MeV)        & 36.976\\
                 & $\hat{r}_0~$(MeV~fm) & 28.170\\
\end{tabular}
\end{ruledtabular}
\end{table}

The pentaquark wave function is a product of four terms: color, flavor, spin and space wave functions. Concerning the color degree-of-freedom, multiquark systems have richer structure than the conventional mesons and baryons. For instance, the $5$-quark wave function must be colorless but the way of reaching this condition can be done through either a color-singlet or a hidden-color channel or both. The authors of  Refs.~\cite{Harvey:1980rva, Vijande:2009kj} assert that it is enough to consider the color singlet channel when all possible excited states of a system are included. However, a more economical way of computing is considering both, the color singlet wave function:
\begin{align}
\label{Color1}
\chi^c_1 &= \frac{1}{\sqrt{18}}(rgb-rbg+gbr-grb+brg-bgr) \times \nonumber \\
&
\times (\bar r r+\bar gg+\bar bb) \,,
\end{align}
and the hidden-color one:
\begin{align}
\label{Color2}
\chi^{c}_k &= \frac{1}{\sqrt{8}}(\chi^k_{3,1}\chi_{2,8}-\chi^k_{3,2}\chi_{2,7}-\chi^k_{3,3}\chi_{2,6}+\chi^k_{3,4}\chi_{2,5} \nonumber \\
& +\chi^k_{3,5}\chi_{2,4}-\chi^k_{3,6}\chi_{2,3}-\chi^k_{3,7}\chi_{2,2}+\chi^k_{3,8}\chi_{2,1}) \,,
\end{align}
where $k=2\,(3)$ is an index which stands for the symmetric (anti-symmetric) configuration of two quarks in the $3$-quark sub-cluster. All color configurations have been used herein, as in the case of the $P_c^+$ hidden-charm pentaquarks studied in Ref.~\cite{Yang:2015bmv}.

In analogy to the study of the $P_c^+$-type bound states in Ref.~\cite{Yang:2015bmv}, we assume that the flavor wave function of the $uudb\bar{b}$ system is composed by $(udb)(\bar{b}u)+(uub)(\bar{b}d)$ and $(uud)(\bar{b}b)$ configurations. According to the SU(2) symmetry in isospin space, the flavor wave functions for the sub-clusters mentioned above are given by:
\begin{align}
B_{11}  &= uub \,, \\
B_{10}  &= \frac{1}{\sqrt{2}}(ud+du)b \,, \\
B_{1-1} &= ddb \,, \\
B_{00} &= \frac{1}{\sqrt{2}}(ud-du)b \,, \\
B_{\frac12,\frac12}^1 &= \frac{1}{\sqrt{6}}(2uud-udu-duu) \,, \\
B_{\frac12,\frac12}^2 &= \frac{1}{\sqrt{2}}(ud-du)u \,, \\
M_{\frac12,\frac12} &= \bar{b}u \,, \\
M_{\frac12,-\frac12} &= \bar{b}d \,, \\
M_{00} &= \bar{b}b \,.
\end{align}
Consequently, the flavor wave-functions for the 5-quark system with isospin $I=1/2$ or $3/2$ are
\begin{align}
&
\chi_{\frac12,\frac12}^{f1}(5) = \sqrt{\frac{2}{3}} B_{11} M_{\frac12,-\frac12} -\sqrt{\frac{1}{3}} B_{10} M_{\frac12,\frac12} \,, \\
&
\chi_{\frac12,\frac12}^{f2}(5) = B_{00} M_{\frac12,\frac12} \,, \\
&
\chi_{\frac12,\frac12}^{f3}(5) = B_{\frac12,\frac12}^1 M_{00} \,, \\
&
\chi_{\frac12,\frac12}^{f4}(5) = B_{\frac12,\frac12}^2 M_{00} \,, \\
&
\chi_{\frac32,\frac32}^{f1}(5) = B_{\frac32,\frac32} M_{00} \,, \\
& 
\chi_{\frac32,\frac32}^{f2}(5) = B_{1,1} M_{\frac12,\frac12} \,,
\end{align}
where the third component of the isospin is set to be equal to the total one without loss of generality because there is no interaction in the Hamiltonian that can distinguish such component.

We consider herein 5-quark bound states with total spin ranging from $1/2$ to $5/2$. Since our Hamiltonian does not have any spin-orbital coupling dependent potential, we can assume that third component of the spin is equal to the total one without loss of generality. Our spin wave function is given by:
\begin{align}
\label{Spin}
\chi_{\frac12,\frac12}^{\sigma 1}(5) &= \sqrt{\frac{1}{6}} \chi_{\frac32,-\frac12}^{\sigma}(3) \chi_{11}^{\sigma}
-\sqrt{\frac{1}{3}} \chi_{\frac32,\frac12}^{\sigma}(3) \chi_{10}^{\sigma} \nonumber \\
&
+\sqrt{\frac{1}{2}} \chi_{\frac32,\frac32}^{\sigma}(3) \chi_{1-1}^{\sigma} \\
\chi_{\frac12,\frac12}^{\sigma 2}(5) &= \sqrt{\frac{1}{3}} \chi_{\frac12,\frac12}^{\sigma 1}(3) \chi_{10}^{\sigma} -\sqrt{\frac{2}{3}} \chi_{\frac12,-\frac12}^{\sigma 1}(3) \chi_{11}^{\sigma} \\
\chi_{\frac12,\frac12}^{\sigma 3}(5) &= \sqrt{\frac{1}{3}} \chi_{\frac12,\frac12}^{\sigma 2}(3) \chi_{10}^{\sigma} - \sqrt{\frac{2}{3}} \chi_{\frac12,-\frac12}^{\sigma 2}(3) \chi_{11}^{\sigma} \\
\chi_{\frac12,\frac12}^{\sigma 4}(5) &= \chi_{\frac12,\frac12}^{\sigma 1}(3) \chi_{00}^{\sigma} \\
\chi_{\frac12,\frac12}^{\sigma 5}(5) &= \chi_{\frac12,\frac12}^{\sigma 2}(3) \chi_{00}^{\sigma}
\end{align}
for $S=1/2$, and
\begin{align}
\chi_{\frac32,\frac32}^{\sigma 1}(5) &= \sqrt{\frac{3}{5}}
\chi_{\frac32,\frac32}^{\sigma}(3) \chi_{10}^{\sigma} -\sqrt{\frac{2}{5}} \chi_{\frac32,\frac12}^{\sigma}(3) \chi_{11}^{\sigma} \\
\chi_{\frac32,\frac32}^{\sigma 2}(5) &= \chi_{\frac32,\frac32}^{\sigma}(3) \chi_{00}^{\sigma} \\
\chi_{\frac32,\frac32}^{\sigma 3}(5) &= \chi_{\frac12,\frac12}^{\sigma 1}(3) \chi_{11}^{\sigma} \\
\chi_{\frac32,\frac32}^{\sigma 4}(5) &= \chi_{\frac12,\frac12}^{\sigma 2}(3) \chi_{11}^{\sigma}
\end{align}
for $S=3/2$, and
\begin{align}
\chi_{\frac52,\frac52}^{\sigma 1}(5) &= \chi_{\frac32,\frac32}^{\sigma}(3) \chi_{11}^{\sigma}
\end{align}
for $S=5/2$. These expressions can be obtained easily considering the 3-quark and quark-antiquark sub-clusters and using SU(2) algebra. They were derived in Ref.~\cite{Yang:2015bmv} for the hidden-charm pentaquarks.

Among the different methods to solve the Schr\"odinger-like 5-body bound state equation, we use the Rayleigh-Ritz variational principle which is one of the most extended tools to solve eigenvalue problems due to its simplicity and flexibility. However, it is of great importance how to choose the basis on which to expand the wave function. The spatial wave function of a $5$-quark system is written as follows:
\begin{equation}
\label{eq:WFexp}
\psi_{LM_L}=\left[ \left[ \left[ \phi_{n_1l_1}(\vec{\rho}\,) \phi_{n_2l_2}(\vec{\lambda}\,)\right]_{l} \phi_{n_3l_3}(\vec{r}\,) \right]_{l^{\prime}} \phi_{n_4l_4}(\vec{R}\,) \right]_{LM_L} \,,
\end{equation}
where the internal Jacobi coordinates are defined as
\begin{align}
\vec{\rho} &= \vec{x}_1-\vec{x}_2 \,, \\
\vec{\lambda} &= \vec{x}_3 - \left( \frac{m_1\vec{x}_1+m_2\vec{x}_2}{m_1+m_2} \right) \,, \\
\vec{r} &= \vec{x}_4 - \vec{x}_5 \,, \\
\vec{R} &= \left( \frac{m_1 \vec{x}_1 + m_2 \vec{x}_2 + m_3 \vec{x}_3}{m_1+m_2+m_3} \right) \nonumber \\
&
- \left( \frac{m_4 \vec{x}_4 + m_5 \vec{x}_5}{m_4+m_5} \right) \,.
\end{align}
This choice is convenient because the center-of-mass kinetic term $T_{CM}$ can be completely eliminated for a nonrelativistic system.

In order to make the calculation tractable, even for complicated interactions, we replace the orbital wave functions, $\phi$'s in Eq.~\eqref{eq:WFexp}, by a superposition of infinitesimally-shifted Gaussians (ISG)~\cite{Hiyama:2003cu}:
\begin{align}
&
\phi_{nlm}(\vec{r}\,) = N_{nl} r^{l} e^{-\nu_{n} r^2} Y_{lm}(\hat{r}) \nonumber \\
&
= N_{nl} \lim_{\varepsilon\to 0} \frac{1}{(\nu_{n}\varepsilon)^l} \sum_{k=1}^{k_{\rm
max}} C_{lm,k} e^{-\nu_{n}(\vec{r}-\varepsilon \vec{D}_{lm,k})^{2}} \,.
\end{align}
where the limit $\varepsilon\to 0$ must be carried out after the matrix elements have been calculated analytically. This new set of basis functions makes the calculation of 5-body matrix elements easier without the laborious Racah algebra~\cite{Hiyama:2003cu}. Moreover, all the advantages of using Gau\ss ians remain with the new basis functions.

Finally, in order to fulfill the Pauli principle, the complete antisymmetric wave-function is written as
\begin{equation}
\label{TPs}
\Psi_{JM,i,j,k,n}={\cal A} \left[ \left[ \psi_{L} \chi^{\sigma_i}_{S}(5) \right]_{JM_J} \chi^{f}_j \chi^{c}_k \right] \,,
\end{equation}
where $\cal{A}$ is the antisymmetry operator of the 5-quark system. This is needed because we have constructed an antisymmetric wave function for only two quarks of the 3-quark sub-cluster, the remaining (anti)-quarks of the system have been added to the wave function by simply considering the appropriate Clebsch-Gordan coefficients. Moreover, the antisymmetry operator $\cal{A}$ has six terms but since we are considering that the $uudb\bar{b}$ system is made by the quark arrangements $(udb)(\bar{b}u)+(uub)(\bar{b}d)$ and $(uud)(\bar{b}b)$, we have 
\begin{equation}
{\cal{A}}_1 = 1-(15)-(25) \,,
\end{equation}
for the $(udb)(\bar{b}u)+(uub)(\bar{b}d)$ configuration, and
\begin{equation}
{\cal{A}}_2 = 1-(13)-(23) \,,
\end{equation}
for the $(uud)(\bar{b}b)$ structure.

%%%%%%%%%%%%%%%%%%%%%%%%%%%%%%%%%%%%%%%%%%%%%%%%%%%%%%%%%%%%%%%%%%%%%%%%%%%%%%%%%%%%%%%%%%

\begin{table*}[!t]
\caption{\label{GCC} All possible channels for hidden-bottom pentaquark systems with negative parity.}
\begin{ruledtabular}
\begin{tabular}{cccccc}
& & \multicolumn{2}{c}{$I=\frac{1}{2}$} & \multicolumn{2}{c}{$I=\frac{3}{2}$} \\[2ex]
$J^P$~~&~Index~ & $\chi_J^{\sigma_i}$;~$\chi_I^{f_j}$;~$\chi_k^c$ & Channel~~ & $\chi_J^{\sigma_i}$;~$\chi_I^{f_j}$;~$\chi_k^c$ & Channel~~ \\
&&$[i; ~j; ~k]$& &$[i; ~j; ~k]$&  \\[2ex]
$\frac{1}{2}^-$ & 1  & $[4,5; ~3,4; ~1]$   & $(N \eta_b)^1$ & $[1; ~1; ~1]$   & $(\Delta \Upsilon)^1$ \\
&  2 & $[4,5; ~3,4; ~2,3]$ & $(N \eta_b)^8$ & $[1; ~1; ~3]$  & $(\Delta \Upsilon)^8$ \\
&  3 & $[2,3; ~3,4; ~1]$   & $(N \Upsilon)^1$  & $[4; ~2; ~1]$   & $(\Sigma_b \bar{B})^1$ \\
&  4 & $[2,3; ~3,4; ~2,3]$ & $(N \Upsilon)^8$ & $[4,5; ~2; ~2,3]$   & $(\Sigma_b \bar{B})^8$   \\
&  5 & $[5; ~2; ~1]$     & $(\Lambda_b \bar{B})^1$  & $[2; ~2; ~1]$   & $(\Sigma_b \bar{B}^*)^1$ \\
&  6 & $[4,5; ~2; ~2,3]$ & $(\Lambda_b \bar{B})^8$  & $[2,3; ~2; ~2,3]$   & $(\Sigma_b \bar{B}^*)^8$ \\
&  7 & $[3; ~2; ~1]$     & $(\Lambda_b \bar{B}^*)^1$  & $[1; ~2; ~1]$   & $(\Sigma^*_b \bar{B}^*)^1$ \\
&  8 & $[2,3; ~2; ~2,3]$ & $(\Lambda_b \bar{B}^*)^8$  & $[1; ~2; ~3]$   & $(\Sigma^*_b \bar{B}^*)^8$ \\
&  9 & $[4; ~1; ~1]$     & $(\Sigma_b \bar{B})^1$     & & \\
& 10 & $[4,5; ~1; ~2,3]$ & $(\Sigma_b \bar{B})^8$     & & \\
& 11 & $[2; ~1; ~1]$     & $(\Sigma_b \bar{B}^*)^1$   & & \\
& 12 & $[2,3; ~1; ~2,3]$ & $(\Sigma_b \bar{B}^*)^8$   & & \\
& 13 & $[1; ~1; ~1]$     & $(\Sigma^*_b \bar{B}^*)^1$ & & \\
& 14 & $[1; ~1; ~3]$     & $(\Sigma^*_b \bar{B}^*)^8$ & & \\[2ex]
$\frac{3}{2}^-$ & 1  & $[3,4; ~3,4; ~1]$   & $(N \Upsilon)^1$ & $[2; ~1; ~1]$   & $(\Delta \eta_b)^1$\\
& 2  & $[3,4; ~3,4; ~2,3]$  & $(N \Upsilon)^8$ & $[2; ~1; ~3]$  & $(\Delta \eta_b)^8$ \\
& 3  & $[4; ~2; ~1]$     & $(\Lambda_b \bar{B}^*)^1$  & $[1; ~1; ~1]$   & $(\Delta \Upsilon)^1$ \\
& 4  & $[3,4; ~2; ~2,3]$  & $(\Lambda_b \bar{B}^*)^8$ & $[1; ~1; ~3]$ & $(\Delta \Upsilon)^8$ \\
& 5  & $[3; ~1; ~1]$     & $(\Sigma_b \bar{B}^*)^1$  & $[3; ~2; ~1]$   & $(\Sigma_b \bar{B}^*)^1$ \\
& 6  & $[3,4; ~1; ~2,3]$  & $(\Sigma_b \bar{B}^*)^8$  & $[3,4; ~2; ~2,3]$   & $(\Sigma_b \bar{B}^*)^8$  \\
& 7  & $[2; ~1; ~1]$     & $(\Sigma^*_b \bar{B})^1$  & $[2; ~2; ~1]$   & $(\Sigma^*_b \bar{B})^1$ \\
& 8  & $[2; ~1; ~3]$  & $(\Sigma^*_b \bar{B})^8$  & $[2; ~2; ~3]$   & $(\Sigma^*_b \bar{B})^8$  \\
& 9  & $[1; ~1; ~1]$   & $(\Sigma^*_b \bar{B}^*)^1$ & $[1; ~2; ~1]$  & $(\Sigma^*_b \bar{B}^*)^1$ \\
& 10 & $[1; ~1; ~3]$  & $(\Sigma^*_b \bar{B}^*)^8$  & $[1; ~2; ~3]$   &  $(\Sigma^*_b \bar{B}^*)^8$  \\[2ex]
$\frac{5}{2}^-$ & 1  & $[1; ~1; ~1]$   & $(\Sigma^*_b \bar{B}^*)^1$ & $[1; ~1; ~1]$   & $(\Delta \Upsilon)^1$ \\
& 2  & $[1; ~1; ~3]$  & $(\Sigma^*_b \bar{B}^*)^8$ & $[1; ~1; ~3]$  & $(\Delta \Upsilon)^8$ \\
& 3  &   &   & $[1; ~2; ~1]$   & $(\Sigma^*_b \bar{B}^*)^1$ \\
& 4  &   & & $[1; ~2; ~3]$ & $(\Sigma^*_b \bar{B}^*)^8$ \\
\end{tabular}
\end{ruledtabular}
\end{table*}

\section{Results}
\label{sec:results}

In the present calculation, we investigate the possible lowest-lying states of the $uudb\bar b$ pentaquark system taking into account the $(udb)(\bar{b}u)+(uub)(\bar{b}d)$ and $(uud)(\bar{b}b)$ configurations in which the considered baryons have always positive parity and the open- and hidden-bottom mesons are either pseudoscalars $(J^P=0^-)$ or vectors $(1^-)$.\footnote{There may exist other baryon-meson structures which contain excited hadrons such as $\chi_{b1}N(940)$, $\Upsilon(1S) N(1440)$ and so on; all of them are beyond the scope of the present calculation.} This means that, in our approach, a pentaquark state with positive parity should have at least one unity of angular momentum: $L=1$, whereas the negative parity states have $L=0$. Reference~\cite{Yang:2015bmv} showed that positive parity $L=1$ hidden-charm pentaquark states are always above its corresponding non-interacting baryon-meson threshold and the same situation is found within the hidden-bottom sector.

For negative parity $L=0$ hidden-bottom pentaquarks we assume that the angular momenta $l_1$, $l_2$, $l_3$, $l_4$, which appear in Eq.~\eqref{eq:WFexp}, are $0$. In this way, the total angular momentum, $J$, coincides with the total spin, $S$, and can take values $1/2$, $3/2$ and $5/2$. The possible baryon-meson channels which are under consideration in the computation are listed in Table~\ref{GCC}, they have been grouped according to total spin and isospin. The third and fifth columns of Table~\ref{GCC} show the necessary basis combination in spin $(\chi^{\sigma_i}_J)$, flavor $(\chi^{f_j}_I)$, and color $(\chi^c_k)$ degrees-of-freedom. The physical channels with color-singlet (labeled with the superindex $1$) and hidden-color (labeled with the superindex $8$) configurations are listed in the fourth and sixth columns.
% Here need to point out that due to the first two identical quarks in three quarks sub-cluster are symmetry ($l=0$) in spatial part, based on the Pauli exclusion principle, it should be  anti-symmetry after coupling the other three spaces for these two quarks.

\begin{table}[!t]
\caption{\label{Gresult1} Lowest-lying states of hidden-bottom pentaquarks with quantum numbers $I(J^P)=\frac12(\frac12^-)$. {\it First column:} channel in which a bound state appears, we show in parenthesis the experimental value, in MeV, of the noninteracting baryon-meson threshold; {\it second column:} color-singlet (S), hidden-color (H) and coupled-channels (S+H) calculation; {\it third column:} theoretical mass, in MeV, of the pentaquark state; {\it fourth column:} its binding energy, in MeV, considering the theoretical baryon-meson threshold; {\it fifth column:} again the pentaquark's mass, in MeV, but re-scaled attending to the experimental baryon-meson threshold. The percentages of color-singlet (S) and hidden-color (H) channels are also given when the coupled-channels calculation is performed. The baryon-meson channels that do not appear here have been also considered in the computation but no bound states were found.}
\begin{ruledtabular}
\begin{tabular}{lcccc}
Channel   & Color & $M$ & $E_B$ & $M'$ \\[2ex]
$\Sigma_b\bar{B}$ & S   & $11080$ & $-15$  & $11074$ \\
$(11089)$         & H   & $11364$ & $+269$ & $11358$ \\
                  & S+H & $11078$ & $-17$  & $11072$ \\
                  & \multicolumn{4}{c}{Percentage (S;H): 98.5\%; 1.5\%}  \\[2ex]
$\Sigma_b\bar{B}^*$ & S   & $11115$ & $-21$  & $11113$ \\
$(11134)$           & H   & $11257$ & $+121$ & $11255$ \\
                    & S+H & $11043$ & $-93$  & $11041$ \\
                    & \multicolumn{4}{c}{Percentage (S;H): 57.9\%; 42.1\%} \\[2ex]
$\Sigma^*_b\bar{B}^*$ & S   & $11127$ & $-26$  & $11128$ \\
$(11154)$             & H   & $10921$ & $-232$ & $10922$ \\
                      & S+H & $10861$ & $-292$ & $10862$ \\
                      & \multicolumn{4}{c}{Percentage (S;H): 15.8\%; 84.2\%} \\ 
%     ~~~~~Mixed (singlet)  & 10195 & &  \\
%     ~~~~~Mixed (full)  & 10195 & & \\  \hline\hline
\end{tabular}
\end{ruledtabular}
\end{table}

\begin{table}[!t]
\caption{\label{Gresult2} Lowest-lying states of hidden-bottom pentaquarks with quantum numbers $I(J^P)=\frac12(\frac32^-)$. {\it First column:} channel in which a bound state appears, we show in parenthesis the experimental value, in MeV, of the noninteracting baryon-meson threshold; {\it second column:} color-singlet (S), hidden-color (H) and coupled-channels (S+H) calculation; {\it third column:} theoretical mass, in MeV, of the pentaquark state; {\it fourth column:} its binding energy, in MeV, considering the theoretical baryon-meson threshold; {\it fifth column:} again the pentaquark's mass, in MeV, but re-scaled attending to the experimental baryon-meson threshold. The percentages of color-singlet (S) and hidden-color (H) channels are also given when the coupled-channels calculation is performed. The baryon-meson channels that do not appear here have been also considered in the computation but no bound states were found.}
\begin{ruledtabular}
\begin{tabular}{lcccc}
Channel   & Color & $M$ & $E_B$ & $M'$ \\[2ex]
$\Sigma_b\bar{B}^*$ & S   & $11124$ & $-12$  & $11122$ \\
$(11134)$           & H   & $11476$ & $+340$ & $11475$ \\
                    & S+H & $11122$ & $-14$  & $11120$ \\
                    & \multicolumn{4}{c}{Percentage (S;H): 99.6\%; 0.4\%} \\[2ex]
$\Sigma^*_b\bar{B}$ & S   & $11097$ & $-15$ & $11094$ \\
$(11109)$           & H   & $11175$ & $+63$ & $11172$ \\
                    & S+H & $11045$ & $-67$ & $11042$ \\
                    & \multicolumn{4}{c}{Percentage (S;H): 55.5\%; 44.5\%} \\[2ex]
$\Sigma^*_b\bar{B}^*$ & S   & $11138$ & $-15$  & $11139$ \\
$(11154)$             & H   & $11051$ & $-102$ & $11052$ \\
                      & S+H & $10958$ & $-195$ & $10959$ \\
                      & \multicolumn{4}{c}{Percentage (S;H): 22.2\%; 77.8\%} \\ 
%
%    ~~~~~Mixed (singlet) & 10246 & & & \\  
%    ~~~~~Mixed (full) & 10246 & & & \\ \hline
\end{tabular}
\end{ruledtabular}
\end{table}

\begin{table}[!t]
\caption{\label{Gresult3} Lowest-lying states of hidden-bottom pentaquarks with quantum numbers $I(J^P)=\frac12(\frac52^-)$. {\it First column:} channel in which a bound state appears, we show in parenthesis the experimental value, in MeV, of the noninteracting baryon-meson threshold; {\it second column:} color-singlet (S), hidden-color (H) and coupled-channels (S+H) calculation; {\it third column:} theoretical mass, in MeV, of the pentaquark state; {\it fourth column:} its binding energy, in MeV, considering the theoretical baryon-meson threshold; {\it fifth column:} again the pentaquark's mass, in MeV, but re-scaled attending to the experimental baryon-meson threshold. The percentages of color-singlet (S) and hidden-color (H) channels are also given when the coupled-channels calculation is performed. The baryon-meson channels that do not appear here have been also considered in the computation but no bound states were found.}
\begin{ruledtabular}
\begin{tabular}{lcccc}
Channel   & Color & $M$ & $E_B$ & $M'$ \\[2ex]
$\Sigma^*_b\bar{B}^*$ & S   & $11141$ & $-12$  & $11151$ \\
$(11154)$             & H   & $11547$ & $+394$ & $11548$ \\
                      & S+H & $11140$ & $-13$  & $11141$ \\
                      & \multicolumn{4}{c}{Percentage (S;H): 99.6\%; 0.4\%} \\
\end{tabular}
\end{ruledtabular}
\end{table}

\begin{table}[!t]
\caption{\label{Gresult4} Lowest-lying states of hidden-bottom pentaquarks with quantum numbers $I(J^P)=\frac32(\frac32^-)$. {\it First column:} channel in which a bound state appears, we show in parenthesis the experimental value, in MeV, of the noninteracting baryon-meson threshold; {\it second column:} color-singlet (S), hidden-color (H) and coupled-channels (S+H) calculation; {\it third column:} theoretical mass, in MeV, of the pentaquark state; {\it fourth column:} its binding energy, in MeV, considering the theoretical baryon-meson threshold; {\it fifth column:} again the pentaquark's mass, in MeV, but re-scaled attending to the experimental baryon-meson threshold. The percentages of color-singlet (S) and hidden-color (H) channels are also given when the coupled-channels calculation is performed. The baryon-meson channels that do not appear here have been also considered in the computation but no bound states were found.}
\begin{ruledtabular}
\begin{tabular}{lcccc}
Channel   & Color & $M$ & $E_B$ & $M'$ \\[2ex]
$\Sigma_b\bar{B}^*$ & S   & $11136$ & $0$    & $11134$ \\
$(11134)$           & H   & $11310$ & $+174$ & $11308$  \\
                    & S+H & $11021$ & $-115$ & $11019$ \\
                    & \multicolumn{4}{c}{Percentage (S;H): 64.7\%; 35.3\%} \\[2ex]
$\Sigma^*_b\bar{B}$ & S   & $11112$ & $0$    & $11109$ \\
$(11109)$           & H   & $11041$ & $-71$  & $11038$ \\
                    & S+H & $10999$ & $-113$ & $10996$ \\
                    & \multicolumn{4}{l}{Percentage (S;H): 18.4\%; 81.6\%} \\[2ex]
$\Sigma^*_b\bar{B}^*$ & S   & $11153$ & $0$    & $11154$ \\
$(11154)$             & H   & $11102$ & $-51$  & $11103$ \\
                      & S+H & $11048$ & $-105$ & $11049$ \\
                      & \multicolumn{4}{c}{Percentage (S;H): 15.7\%; 84.3\%} \\ 
%    mixed (only color-singlet) & 10690 & & & \\  
%    mixed (full) & 10690 & & & \\  \hline
\end{tabular}
\end{ruledtabular}
\end{table}

\begin{table}[!t]
\caption{\label{Gresult5} Lowest-lying states of hidden-bottom pentaquarks with quantum numbers $I(J^P)=\frac32(\frac52^-)$. {\it First column:} channel in which a bound state appears, we show in parenthesis the experimental value, in MeV, of the noninteracting baryon-meson threshold; {\it second column:} color-singlet (S), hidden-color (H) and coupled-channels (S+H) calculation; {\it third column:} theoretical mass, in MeV, of the pentaquark state; {\it fourth column:} its binding energy, in MeV, considering the theoretical baryon-meson threshold; {\it fifth column:} again the pentaquark's mass, in MeV, but re-scaled attending to the experimental baryon-meson threshold. The percentages of color-singlet (S) and hidden-color (H) channels are also given when the coupled-channels calculation is performed. The baryon-meson channels that do not appear here have been also considered in the computation but no bound states were found.}
\begin{ruledtabular}
\begin{tabular}{lcccc}
Channel   & Color & $M$ & $E_B$ & $M'$ \\[2ex]
$\Sigma^*_b\bar{B}^*$ & S   & $11052$ & $-101$ & $11053$ \\
$(11154)$             & H   & $10974$ & $-179$ & $10975$ \\
                      & S+H & $10931$ & $-222$ & $10932$ \\
                      & \multicolumn{4}{c}{Percentage (S;H): 19.9\%; 80.1\%} \\ 
%    Mixed (singlet) & 10741 & & & \\  
%    Mixed (full) & 10741 & & & \\
\end{tabular}
\end{ruledtabular}
\end{table}

\begin{table}[!t]
\caption{\label{tab:disqq} The distance, in fm, between any two quarks of the found pentaquark bound-states.}
\begin{ruledtabular}
\begin{tabular}{ccccccc}
$I(J^{P})$ & Channel & Mixing & $r_{qq}$ & $r_{qQ}$ & $r_{q\bar{Q}}$ & $r_{Q\bar{Q}}$  \\[2ex]
$\frac12({\frac{1}{2}}^{-})$
& $\Sigma_b \bar{B}$     & S   & 1.17 & 0.87 & 1.02 & 1.00 \\
&                        & S+H & 1.13 & 0.84 & 0.98 & 0.94 \\
& $\Sigma_b \bar{B}^*$   & S   & 1.09 & 0.81 & 0.92 & 0.82 \\
&                        & S+H & 0.94 & 0.70 & 0.71 & 0.34 \\
& $\Sigma_b^* \bar{B}^*$ & S   & 1.06 & 0.79 & 0.88 & 0.75 \\
&                        & S+H & 0.91 & 0.71 & 0.70 & 0.24 \\[2ex]
$\frac12({\frac{3}{2}}^{-})$
& $\Sigma_b \bar{B}^*$   & S   & 1.23 & 0.90 & 1.09 & 1.09 \\
&                        & S+H & 1.21 & 0.90 & 1.07 & 1.07 \\
& $\Sigma_b^* \bar{B}$   & S   & 1.18 & 0.88 & 1.04 & 1.01 \\
&                        & S+H & 0.98 & 0.74 & 0.74 & 0.34 \\
& $\Sigma_b^* \bar{B}^*$ & S   & 1.17 & 0.87 & 1.02 & 0.97 \\
&                        & S+H & 0.95 & 0.72 & 0.72 & 0.25 \\[2ex]
$\frac12({\frac{5}{2}}^{-})$
& $\Sigma_b^* \bar{B}^*$ & S   & 1.25  & 0.92 & 1.11 & 1.13 \\
&                        & S+H &1.25  & 0.92 & 1.11 & 1.11 \\[2ex]
$\frac32({\frac{3}{2}}^{-})$ 
& $\Sigma_b \bar{B}^*$   & S+H & 1.02 & 0.78  & 0.77 & 0.27 \\
& $\Sigma_b^* \bar{B}$   & S+H & 1.02 & 0.84  & 0.82 & 0.26 \\
& $\Sigma_b^* \bar{B}^*$ & S+H & 1.05 & 0.83  & 0.81 & 0.26 \\[2ex]
$\frac32({\frac{5}{2}}^{-})$ 
& $\Sigma_b^* \bar{B}^*$ & S   & 1.03 & 0.86 & 0.86 & 0.29 \\                             
&                        & S+H & 1.00 & 0.86 & 0.84 & 0.26 \\
\end{tabular}
\end{ruledtabular}
\end{table}

Tables ranging from~\ref{Gresult1} to~\ref{Gresult5} summarize our findings about the possible existence of lowest-lying hidden-bottom pentaquarks with quantum numbers $I(J^P)=\frac{1}{2}(\frac{1}{2}^-)$, $\frac{1}{2}(\frac{3}{2}^-)$, $\frac{1}{2}(\frac{5}{2}^-)$, $\frac{3}{2}(\frac{3}{2}^-)$ and $\frac{3}{2}(\frac{5}{2}^-)$, respectively.\footnote{A table associated with the $I(J^P)=\frac{3}{2}(\frac{1}{2}^-)$ sector is not shown because no bound-states were found.} In each Table, the first column shows the baryon-meson channel in which a bound state appears, it also indicates in parenthesis the experimental value of the noninteracting baryon-meson threshold; the second column refers to color-singlet (S), hidden-color (H) and coupled-channels (S+H) calculations; the third and fourth columns show the theoretical mass and binding energy of the pentaquark bound-state; and the fifth column presents the theoretical mass of the pentaquark state but re-scaled attending to the experimental baryon-meson threshold, this is in order to avoid theoretical uncertainties coming from the quark model prediction of the baryon and meson spectra. The percentages of color-singlet (S) and hidden-color (H) channels are also given when the coupled-channels calculation is performed. For the channels in which a bound-state is found, we show in Table~\ref{tab:disqq} a calculation of all possible quark-quark distances in order to get some insight about the kind of pentaquark we are dealing with: molecular or compact. 

We proceed now to describe in detail our theoretical findings:

{\bf The $\bm{I(J^P)=\frac12(\frac12^-)}$ channel:} Among all the possible baryon-meson channels: $N\eta_b$, $N\Upsilon$, $\Lambda_b \bar{B}$, $\Lambda_b \bar{B}^*$, $\Sigma_b \bar{B}$, $\Sigma_b \bar{B}^*$ and $\Sigma_b^* \bar{B}^*$; only the last three point to the possibility of having bound states. In particular, when considering only the color-singlet configuration of $\Sigma_b \bar{B}$, $\Sigma_b \bar{B}^*$ and $\Sigma_b^* \bar{B}^*$ the binding energies are $-15\,\text{MeV}$, $-21\,\text{MeV}$ and $-26\,\text{MeV}$, respectively. This motivates the possibility of finding molecular-type baryon-meson structures around the $\Sigma_b^{(\ast)}\bar B^{(\ast)}$ thresholds. One can see in Table~\ref{Gresult1} that the binding energy is slightly larger for $\Sigma_b \bar B$ $(E_B=-17\,\text{MeV})$ when the hidden-color configuration is incorporated in the calculation; in fact, its contribution to the hadron's wave function is pretty small, $1.5\%$. This small change could indicate that the state is of molecular-type and, in fact, all interquark distances shown in Table~\ref{tab:disqq} are very similar, $\sim\!\! 1\,\text{fm}$, pointing to a relatively extended hadron. The situation is quite different for the other two bound states found in the $\Sigma_b \bar{B}^*$ and $\Sigma_b^* \bar{B}^*$ channels. One can see in Table~\ref{Gresult1} that the binding energy becomes very large when the hidden-color configuration is incorporated: $E_B=-93\,\text{MeV}$ for $\Sigma_b \bar{B}^*$ and $E_B=-292\,\text{MeV}$ for $\Sigma_b^* \bar{B}^*$. These deeply bound states are usually associated with compact multiquark structures. One can observe in Table~\ref{tab:disqq} that the distance between the two heavy quarks reduces considerably when the hidden-color configuration is incorporated, indicating that a compact heavy quark-antiquark core is formed and surrounded by light quarks. In comparison with our study of hidden-charm pentaquarks of Ref.~\cite{Yang:2015bmv}, a similar trend is observed but, as expected, we find in the hidden-bottom sector larger binding energies and bigger contributions of the hidden-color configurations. As an example of the last feature, we have $42.1\%$ for $(\Sigma_b \bar{B}^*)^8$ and $84.2\%$ for $(\Sigma_b^* \bar{B}^*)^8$ which compare with $32.6\%$ for $(\Sigma_c \bar{D}^*)^8$ and $77\%$ for $(\Sigma^*_c \bar{D}^*)^8$.

%%%%%%%%%%

{\bf The $\bm{I(J^P)=\frac12(\frac32^-)}$ channel:} There exist bound-states in the  $\Sigma_b \bar{B}^*$, $\Sigma_b^* \bar{B}$ and $\Sigma_b^* \bar{B}^*$ configurations but no signal of binding is found for the baryon-meson channels $N\Upsilon$ and $\Lambda_b \bar{B}^*$. Looking at Table~\ref{Gresult2}, one can realize that the binding energies are $-12\,\text{MeV}$, $-15\,\text{MeV}$ and $-15\,\text{MeV}$ for the $\Sigma_b \bar{B}^*$, $\Sigma_b^* \bar{B}$ and $\Sigma_b^* \bar{B}^*$ channels, respectively, when considering only the color-singlet configuration. This motivates again the possibility of finding molecular-type baryon-meson structures around the $\Sigma_b^{(\ast)}\bar B^{(\ast)}$ thresholds. If one incorporates in the coupled-channels calculation the hidden-color configurations, the situation in the $I(J^P)=\frac12(\frac32^-)$ channel is quite similar with respect the $I(J^P)=\frac12(\frac12^-)$ one. While the $\Sigma_b \bar B^*$ bound-state modifies slightly its mass and points to a molecular-type structure (see Table~\ref{tab:disqq} for comparing interquark distances) the other two states found in $\Sigma_b^* \bar{B}$ and $\Sigma_b^* \bar{B}^*$ channels appear to be tightly bound with binding energies $-67\,\text{MeV}$ and $-195\,\text{MeV}$, respectively. One can see in Table~\ref{tab:disqq} that, for the deeply-bound states, the distance between the heavy quark-antiquark pair reduces considerably and the other interquark distances, despite becoming smaller, are much larger and of the same order of magnitude. This could point to a possible compact multiquark nature of these states as explained above. It is also interesting to mention herein that the contribution to the wave function of the hidden color configuration is negligible for the $\Sigma_b \bar{B}^*$ channel, slightly sub-dominant in the $\Sigma_b^* \bar{B}$ case and about $80\%$ for the $\Sigma_b^* \bar{B}^*$ channel. In Ref.~\cite{Yang:2015bmv}, we assigned to the $P_c(4380)^+$ signal observed by the LHCb Collaboration~\cite{Aaij:2015tga} a bound state found in the $I(J^P)=\frac12(\frac32^-)$ $\Sigma^*_c \bar{D}$ channel. Its hidden-bottom pentaquark partner would be the bound state in the $\Sigma_b^*\bar B$ channel shown in Table~\ref{Gresult2}. Both states present similar characteristics beyond the typical differences associated with having different heavy quark content. The quantum numbers assignment of the $P_c(4450)^+$ signal is not yet clear and different possibilities are currently discussed in the literature~\cite{Chen:2016qju}, this reason avoids us to comment herein on what would be the hidden-bottom pentaquark partner of this state.

%%%%%%%%%%

{\bf The $\bm{I(J^P)=\frac12(\frac52^-)}$ channel:} The $\Sigma_b^* \bar{B}^*$ channel is the only baryon-meson structure needed to be considered (see Table~\ref{GCC}). When the computation is performed taking into account only the singlet-color configuration, a slightly bound state is found with a binding energy of $-12\,\text{MeV}$ (see now Table~\ref{Gresult3}). If the hidden-color channel is incorporated to the coupled-channels calculation the binding energy increases just by $1\,\text{MeV}$. This indicates that the effect of this channel is very small, i.e. the contribution of the hidden-color configuration to the bound-state wave function is negligible, $0.4\%$. As one can see in Table~\ref{tab:disqq}, the distance between any pair of quarks is around $1\,\text{fm}$ which is a common feature within our framework for all bound states dominated by the color-singlet configuration. Again, we remark herein that this peculiarity could point to have a state of molecular nature.

%%%%%%%%%%

{\bf The $\bm{I(J^P)=\frac32(\frac32^-)}$ channel:} Table~\ref{Gresult4} indicates that no bound states are found in any baryon-meson configuration when color-singlet arrangements are the only ones considered in the coupled-channels calculation. If hidden-color clusters are added to the computation, a bound state appears in the $\Sigma_b\bar{B}^\ast$, $\Sigma_b^\ast\bar{B}$ and $\Sigma_b^\ast\bar{B}^\ast$ channels. The bound state found in the $\Sigma_b\bar{B}^\ast$ channel is quite sensitive to the numerical set-up, making our prediction not very trustable. However, the other two bound-states are very stable against numerical checks and also reflect similar features: (i) the $(\Sigma_b^\ast\bar{B})$- and $(\Sigma_b^\ast\bar{B}^\ast)$-type states are deeply bound with binding energies around $100\,\text{MeV}$, (ii) the hidden-color configurations contribute $\sim\!80\%$ to the wave function, and (iii) the distance of the $b\bar b$-pair is much smaller than the others (see Table~\ref{tab:disqq}) indicating that there is a quark-antiquark core surrounded by three light quarks.

%%%%%%%%%%

{\bf $\bm{I(J^P)=\frac32(\frac52^-)}$ channel:} Only two baryon-meson channels contribute to this case: $\Delta\Upsilon(1S)$ and $\Sigma_b^\ast \bar{B}^\ast$. As in all cases studied before, we do not find any bound state in the $\Delta\Upsilon(1S)$ configuration. However, a bound-state is found in the $\Sigma_b^\ast \bar{B}^\ast$ channel when considering either the singlet- or hidden-color configurations; the coupling between them just increases the binding energy of the state. We can see in Table~\ref{Gresult5} that, in the complete coupled-channels calculation, the bound state has a binding energy of around $200\,\text{MeV}$ and the singlet-color configuration is subdominant, contributing $20\%$ to the formation of the hadron. Table~\ref{tab:disqq} reflects again that this state is a heavy quark-antiquark core surrounded by light quarks.

%%%%%%%%%%%%%%%%%%%%%%%%%%%%%%%%%%%%%%%%%%%%%%%%%%%%%%%%%%%%%%%%%%%%%%%%%%%%%%%%%%%%%%%%%%

\section{Epilogue}
\label{sec:summary}

The $P_c(4380)^+$ and $P_c(4450)^+$ structures were discovered by the LHCb Collaboration in 2015. They have capture the interest of many theorists because their possible hidden-charm pentaquark composition since they were observed in the $J/\psi p$ invariant mass spectrum of the $\Lambda^{0}_{b}\to J/\psi K^{-}p$ decay. The measurement of hadrons containing valence $c$-quarks has been historically followed by the identification of similar structures with $b$-quark content. Therefore, it is reasonable to expect a subsequent observation of the bottom analogues of the $P_c(4380)^+$ and $P_c(4450)^+$ resonances.

In Ref.~\cite{Yang:2015bmv}, within a chiral quark model formalism, the $P_c(4380)^+$ was suggested to be a bound state of $\Sigma_c^\ast\bar{D}$ with quantum numbers $J^P=\frac{3}{2}^-$. The nature of the $P_c(4450)^+$ signal was not clearly established because, despite of having a couple of possible candidates attending to the agreement between theoretical and experimental masses, there was an inconsistency between the parity of the state determined experimentally and those predicted theoretically. Further pentaquark bound-states which contain dominant $\Sigma_c\bar{D}$ and $\Sigma_c^\ast\bar{D}^\ast$ Fock-state components were also found in the region about $4.3-4.5\,\text{GeV}$.

The work presented herein constitutes a natural extension of the analysis performed in Ref.~\cite{Yang:2015bmv}. We have studied the possibility of having pentaquark bound-states in the hidden-bottom sector with quantum numbers $J^P=\frac12^\pm$, $\frac32^\pm$ and $\frac52^\pm$, and in the $\frac12$ and $\frac32$ isospin sectors. The chiral quark model used is based on the existence of Goldstone-boson exchange interactions between light quarks that are encoded in a phenomenological potential which already contains the perturbative one-gluon exchange and the nonperturbative linear-screened confining terms. Note also that the model parameters have been fitted in the past through hadron, hadron-hadron and multiquark phenomenology. Moreover, the five-body bound state problem is solved by means of the Gau\ss ian expansion method which allows us to compute straightforwardly the different matrix elements and it is as accurate as a Faddeev calculation.

We have not found any positive parity hidden-bottom pentaquark state within the scanned quantum numbers: $J=\frac12$, $\frac32$, $\frac52$ and $I=\frac12$, $\frac32$. However, several hidden-bottom pentaquark bound states with negative parity have been identified. These are characterized by the following features: (i) bottom-baryon$+$open-bottom meson such as $\Sigma_b^{(\ast)}\bar{B}^{(\ast)}$ configurations are the dominant ones, (ii) molecular-type bound-states are favored when only color-singlet arrangements are considered in the coupled-channels calculation, (iii) structures in which a compact $b\bar b$-pair is surrounded by three light quarks appear frequently when hidden-color configurations are added to the calculation, (iv) slightly bound states are found when the singlet-color configuration dominates over the hidden-color one whereas deeply bound states appear when the roles of the color configurations are reversed.

It is worth to highlight here that the hidden-bottom pentaquark partner of the $P_c(4380)^+$ signal observed by the LHCb Collaboration would be a bound state in the $\Sigma_b^*\bar B$ channel with quantum numbers $I(J^P)=\frac12(\frac32^-)$ and a mass around $11.04-11.09\,\text{GeV}$. In the complete coupled-channels calculation, both singlet- and hidden-color configurations play an important role contributing almost equally to the formation of the state, $55\%$ and $45\%$ respectively. We have avoided to comment on what would be the hidden-bottom pentaquark partner of the $P_c(4450)^+$ signal because the quantum numbers assignment of this state is still under discussion.

%Finally, our findings resemble the ones found in the hidden-charm sector but, as expected, we find in the hidden-bottom sector larger binding energies and bigger contributions of the hidden-color configurations.

%%%%%%%%%%%%%%%%%%%%%%%%%%%%%%%%%%%%%%%%%%%%%%%%%%%%%%%%%%%%%%%%%%%%%%%%%%%%%%%%%%%%%%%%%%

\begin{acknowledgments}
G.Y. would like to thank L. He for his support and informative discussions.
Work partially financed by: National Natural Science Foundation of China under Grant nos. 11535005 and 11775118; European Union's Horizon 2020 research and innovation programme under the Marie Sk\l{}odowska-Curie grant agreement no. 665919; Spanish MINECO's Juan de la Cierva-Incorporaci\'on programme with grant agreement no. IJCI-2016-30028; and by Spanish Ministerio de Econom\'ia, Industria y Competitividad under contract nos. FPA2014-55613-P, FPA2017-86989-P and SEV-2016-0588.
\end{acknowledgments}

%%%%%%%%%%%%%%%%%%%%%%%%%%%%%%%%%%%%%%%%%%%%%%%%%%%%%%%%%%%%%%%%%%%%%%%%%%%%%%%%%%%%%%%%%%

\bibliography{PentaquarksBottomonium}

\end{document}